\documentclass[prl,twocolumn,nofootinbib,showkeys,showpacs]{revtex4-1}
\usepackage{amssymb}
\usepackage{amsmath}
\usepackage{amstext}
\usepackage{amsfonts}
\usepackage{bbold}
\usepackage{bm}
\usepackage{graphics}
\usepackage{mathtools}
\usepackage{lmodern}
\usepackage{graphicx}
\usepackage[capitalise]{cleveref}

\newcommand{\be}{\begin{equation}}
\newcommand{\ee}{\end{equation}}

\newcommand{\sfrac}[2]{{\textstyle\frac{#1}{#2}}} 

\begin{document}

\title{Lorentz from Galilei, deductively}

\author{Alberto~Nicolis}
\affiliation{Department of Physics, Center for Theoretical Physics, Columbia University, 538W 120th Street, New York, NY, 10027, USA}

\begin{abstract}
I argue that in the Lagrangian formulation of standard, Galilei-invariant Newtonian mechanics there are subtle but concrete signs of {\em Lorentz} invariance. In fact, in a specific sense made explicit in the paper, Newtonian mechanics is {more} Lorentz-invariant than Galilei-invariant. So, special relativity could have been discovered deductively, before there were any indications---such as Maxwell's equations---that Galilei relativity had to be modified. To make this anti-historical exercise less academic, I derive certain velocity-dependent corrections to  long-range interactions between spinless point particles. Such corrections are universal; in particular, they do not depend on the spin of the field mediating such interactions or on how strongly coupled such a field is. I discuss potential applications to the post-Newtonian expansion of general relativity.

\end{abstract}

\maketitle


\noindent\emph{\textbf{Introduction}} --- I want to rediscover special relativity. My only excuse to embark in such a foolish enterprise (cf.~\cite{borges}) is that, as I will show,  one can do so just by staring long enough at Newtonian mechanics. To minimize the number of conditional clauses, and for the sake of seeing how far one can push the argument, in the next two sections I will pretend that we are back in a pre-Maxwell equations era, that we know nothing about special relativity, and that we have no indications that Newtonian physics has to be corrected. I will only use the Lagrangian formulation of Newtonian mechanics, and minimal input from some general (admittedly anachronistic) ideas of effective theories. I will argue that Galilei invariance hints at a symmetry that transforms {\em time} as well. Taking this hint seriously, I will derive Lorentz invariance as the only physically sensible possibility, up to the value (and sign) of $c^2$. I will then apply this logic to potential interactions between spinless particles, and find that these have to be corrected by certain universal velocity-dependent terms. 

I will eventually come back to the real world, and perform a number of nontrivial checks on the results derived.
For now, I  ask the reader to forget everything they know about special relativity.

%
%
\vspace{0.5em}

\noindent\emph{\textbf{A new symmetry of space and time?}} --- Consider the equation of motion for a free  particle of mass $m$:
\be \label{eom}
m \, \ddot {\vec x} (t) = 0 \; .
\ee
It is invariant under translations, rotations, and Galilean boosts. The last act as
\be \label{boost}
\vec x(t) \to \vec x \, '(t) = \vec x(t) - \vec v_0 \, t \; ,
\ee
where $\vec v_0$ is a constant vector, parametrizing the relative velocity between two different inertial frames. For simplicity, in the following I will restrict to infinitesimal $\vec v_0$'s. 

The equation of motion above can be derived from the action
\be \label{action}
S [\vec x \, ] = \int dt \, L[\vec x \, ](t) \; , \qquad L[\vec x \, ] = \sfrac12 m \,  \dot {\vec x} \, ^2 \; ,
\ee
which is a functional of the dynamical variable $\vec x(t)$.
Such an action is manifestly invariant under translations and rotations, but under the Galilean boost \eqref{boost} it is invariant only up to a total time derivative term: to first order in $\vec v_0$,
\be \label{transformed}
S[\vec x \, ] \to S[\vec x \, '] = S[\vec x \, ] - \int dt \, \frac{d}{dt} \big(m \vec v_0 \cdot \vec x(t) \big) \; .
\ee
Such a total derivative term does not affect the variational principle (which involves variations of $\vec x(t)$ that vanish at the boundary) and thus the equation of motion, and so this could be the end of the story. However, it is interesting to ponder whether in fact such a total derivative might be hinting at something deeper.

To see why that could be the case, consider a more familiar example: invariance under standard (infinitesimal) time translations. If we define these as acting on the dynamical variable only,
\be \label{active}
\vec x(t) \to \vec x \, '(t) = \vec x(t - \epsilon) \simeq  \vec x(t) - \epsilon \, \dot {\vec x}(t) \; ,
\ee
and we don't change the time coordinate in the action (the so-called active viewpoint), the action \eqref{action} changes by a boundary term:
\be \label{time translated}
S[\vec x \, ] \to S[\vec x \, '] = S[\vec x \, ] - \int dt \,  \frac{d}{dt} \big (\epsilon \,  \sfrac12 m \, \dot {\vec x} \, ^2  \big) \; .
\ee
However, if we take the equivalent (so-called passive) viewpoint on time-translations as  a redefinition of the time coordinate by a constant shift and a corresponding transformation of the dynamical variable, stating that this takes the same value at the same physical time, regardless of how we label time,
\begin{align}
& t \to t' = t + \epsilon   \label{active t}\\
& \vec x(t) \to \vec x \, '(t') = \vec x (t) \; , \label{active x}
\end{align}
then we get the more standard statement that the action \eqref{action} is invariant.

One way to see this, which is convenient for what follows, is to consider the infinitesimal contribution to the action in a time interval $dt$, $dS = dt L$. Given eqs.~\eqref{active t} and \eqref{active x} above, clearly we have
\be \label{dS}
dt' L[\vec x \,'](t') = dt \, L[\vec x \,](t)
\ee
(if $L$ depended explicitly on time, say through some nontrivial $m(t)$, this would not be true, and in fact in that case we would say that time-translations are not a symmetry of the system.)

From here, we can easily recover the transformation properties of $L$ under an {\em active} transformation, eq.~\eqref{time translated}: to first order in $\epsilon$,
\begin{align}
dt L[\vec x \, ](t) &=  dt' L[\vec x \,'](t' = t + \epsilon) \\
& \simeq dt \Big[ L[\vec x \,'](t) + \epsilon \frac{d}{dt}L[\vec x \, ](t) \Big] \; ,
\end{align}
which is just a rewriting of \eqref{time translated}.
%

This simple example shows that if an action varies by a total derivative term under an active symmetry transformation, it could be a sign that the symmetry under consideration secretly involves a redefinition of the coordinates on which the dynamical variables depend---time, in our case\footnote{Readers from the future familiar with general relativity can consider an analogous phenomenon: an action such as $S = \int d^4 x \sqrt{g} \, {\cal L}$, where ${\cal L}$ is any scalar Lagrangian density, is invariant under diffeomorphisms, that is, arbitrary changes of coordinates. However, if we interpret these as active gauge transformations, 
\be
g_{\mu\nu} \to g_{\mu\nu} - \nabla_\mu \xi_\nu - \nabla_\nu \xi_\mu \; , \quad {\cal L} \to {\cal L} - \xi^\mu \partial_\mu {\cal L} \; ,
\ee 
without ever changing coordinates, then the action changes by a boundary term: $S \to S -   \int d^4 x \sqrt{g} \, \nabla_\mu (\xi^\mu \cal L)$.}. 
So, the natural question is: is there a redefinition of time secretely associated with the boost transformation \eqref{boost}, such that if one takes that into account, our action is invariant? To answer this question, first we must understand under what general conditions a boundary term like that in \eqref{transformed} can be reabsorbed by a redefinition of time. Given the considerations of the last paragraph, the answer is quite simple. 

Consider a general action for $\vec x(t)$, and suppose that, in the sense of \eqref{dS}, the infinitesimal action element is invariant under a certain infinitesimal redefinition of the time variable
\be
t' = t + \xi(t) 
\ee
and a corresponding infinitesimal transformation of the dynamical variable, $\vec x(t) \to \vec x \, ' (t')$. Then, from \eqref{dS}, to first order in the transformation parameters we have
\begin{align}
dt L[\vec x \, ](t) &=  dt' L[\vec x \,'](t' ) \\
& \simeq dt \Big[ L[\vec x \,']+ \frac{d}{dt} \xi L[\vec x \,] +  \xi \frac{d}{dt}L[\vec x \,] \Big] (t) \\
& = dt \Big[ L[\vec x \,'] + \frac{d}{dt} \big( \xi L[\vec x \,] \big)  \Big]  (t) \; .
\end{align}
That is, from the active viewpoint (no change of time coordinate),
\be \label{active}
L[\vec x \,'] (t) = L[\vec x \, ](t) - \frac{d}{dt} \big( \xi L[\vec x ] \big)  (t) \; .
\ee
Notice that, in principle, the time-dependence of $\xi$, like that of $L$, can also come from $\vec x(t)$. That is, for any given trajectory $\vec x(t)$, we can decide to perform a change of time variable that depends on $\vec x(t)$ itself. This possibility is relevant for what follows, and the above manipulations apply unaltered.

%

So, we see that  a total derivative term resulting from an active symmetry transformation,
\be
L[\vec x \,'](t) = L[\vec x \,](t) + \frac{d}{dt}F[\vec  x \, ](t) \; ,
\ee 
is reabsorbable by a suitable redefinition of time if and only if there exists a $\xi(t)$ such that, up to an integration constant,
\be \label{F}
F(t) = - \xi (t ) L(t) 
\ee
(I am now displaying the time-dependence only, which can in part be explicit, and in part come from $\vec x(t)$).
At face value, this is an empty statement: for given $F$ and $L$, it is enough to choose $\xi = - F/L$. However, physical considerations  make this a very non-trivial requirement: if $F$ is not manifestly proportional to the Lagrangian, then every time we modify the Lagrangian of the system, say by adding more particles and including interactions among them, we have to adjust our symmetry transformation---in uglier and uglier ways---in order to keep the action invariant. A symmetry with this property hardly sounds like a fundamental physical symmetry of Nature. And even if it were, it wouldn't be of much use: we would need to know the masses, positions, and interactions of all particles in the universe in order to figure out how time transforms under the symmetry.

We can now come back to the case we are interested in: invariance of our action \eqref{action} under the boost transformations \eqref{boost}. In fact, since we are being picky about total derivative terms, we have to ask whether our action is the most general one yielding the equation of motion \eqref{eom} and with the right symmetries. It is not. Without giving up manifest translational and rotational invariance, the most general one compatible with up-to-total-derivatives boost invariance \eqref{boost} is
\be \label{new action}
S[\vec x \, ] = \int dt \,\big[\sfrac12 m \dot {\vec x} \, ^2  - V_0 \big] \; ,
\ee
where $V_0$ is a constant potential energy, whose value and sign are, at this stage, completely arbitrary.

As a side comment, notice that if one does not insist on the eom \eqref{eom} but just on the symmetries, one could in principle also include terms that depend on the acceleration and higher derivatives of $\vec x$ (which would contribute third- and higher-derivative terms to the eom). However, following standard effective theory logic\footnote{See the future.}, these will be multiplied by coefficients involving inverse powers of some high frequency $\omega_0$---for instance, for a solid body, one does expect such terms, with $\omega_0$ given by the frequency of the lowest-lying vibrational mode. Such terms can thus be neglected for accelerations that are small enough, schematically $a \ll \omega_0 v$.

Under the active boost transformation \eqref{boost}, the new action \eqref{new action} still transforms as in eq.~\eqref{transformed}. Thus, we have
\be
F(t) = - m \vec v_0 \cdot \vec x(t) \; .
\ee
This does not look proportional to the Lagrangian in \eqref{new action}, and so it appears that we have failed. However, if we assume that $V_0$ is nonzero, something interesting happens at very low particle speeds: the Lagrangian is approximately constant, $L \simeq -V_0$, $F(t)$ is thus trivially proportional to it, and the corresponding transformation of time that makes the action invariant under Galilean boosts is
\be
t' \simeq t - \frac{m}{V_0} \, \vec v_0 \cdot \vec x  \; , \qquad (\dot {\vec x} \,^2 \ll V_0/m) \; .
\ee
This, combined with the transformation of $\vec x$ (eq.~\eqref{boost}), is a beautiful symmetry: it can be thought of as a generalization of ordinary rotations, but now acting in space {and} time. 

To see this, notice that $V_0$ has units of energy, and so 
\be 
c^2 \equiv V_0 /m 
\ee
has units of a velocity squared. We can momentarily use $c \equiv \sqrt{|c^2|}$ to convert the units of $t$ to those of $\vec x$. Then, depending on the sign of $c^2$, our transformation laws simply read
\begin{align}
(c t') & = (c t) + \frac{\vec v_0}{c} \cdot \vec x    \qquad (c^2<0)    \\
\vec x \, ' & = \vec x - \frac{\vec v_0}{c}  (c t) 
\end{align}
or 
\begin{align}
(c t') & = (c t) - \frac{\vec v_0}{c} \cdot \vec x    \qquad (c^2>0)    \\
\vec x \, ' & = \vec x - \frac{\vec v_0}{c}  (c t) 
\end{align}

The former case (negative $c^2$) literally describes infinitesimal rotations of angle $v_0/c$ acting in the ``plane" spanned by $ct$ and $\hat v_0 \cdot \vec x$. Together with ordinary spatial rotations, such rotations close the algebra of $SO(4)$---rotations in a four-dimensional space---and in such a case time has no special status compared to the spatial coordinates. The latter case (positive $c^2$) describes some sort of hyperbolic rotations, still acting in the same ``plane". Together with ordinary spatial rotations, they close the algebra of $SO(3,1)$.

To summarize: at generic speeds, our action \eqref{new action} is invariant under Galilei boosts \eqref{boost} only up to total derivatives; however, at very low speeds ($\dot x \ll c$) it becomes strictly invariant under the transformations above, which can be interpreted as geometric transformations acting in space-{\em time}, closely analogous to ordinary spatial rotations.

This could  just be a curiosity. However, suppose that our constant speed parameter $c$ is so large that no experiment has been able to probe physics at speeds close to $c$. Then, instead of interpreting the geometric space-time transformations above as an approximate, low-speed symmetry of our fundamental action \eqref{new action}, we can ponder whether in fact they define a fundamental geometric symmetry of space-time, and interpret our action \eqref{new action} as a low-speed approximation to a more fundamental action that is invariant under such a symmetry. At speeds much smaller than $c$, both possibilities are consistent with what we have been discussing so far.

If we follow the more speculative possibility, first we have to modify the form of the active boost transformation \eqref{boost}. This is because for the transformations above to have a geometric interpretation, we must demand that, for a particle with trajectory $\vec x(t)$, the transformed trajectory $\vec x \, '(t')$ be just the space-time `rotated' version of $\vec x(t)$:
\begin{align}
& t' = t - \frac{1}{c^2} \vec v_0 \cdot \vec x \label{time new}\\
& \vec x \, '(t') = \vec x(t) - \vec v_0 t \; . \label{space new}
\end{align}
So, eq.~\eqref{boost} has to be modified as
\be \label{new boost}
\vec x(t) \to \vec x \, '(t) = \vec x(t) - \vec v_0 \, t  +  \frac{1}{c^2} \dot {\vec x}  \, \vec v_0 \cdot \vec x   \; ,
\ee
This reduces precisely to eq.~\eqref{boost} at low speeds compared to $c$, and so our low-speed analysis above is still valid.

Then, we want to find the fundamental, invariant action of which \eqref{new action} is supposed to be the low-speed approximation.
For the action to be strictly invariant under our putative symmetry \eqref{time new}, \eqref{space new}, we must still obey eq.~\eqref{active}. That is, under an active transformation we must have
\be \label{Lorentz}
L[\vec x \,'] = L[\vec x \,] + \frac{1}{c^2}\frac{d}{dt} \big(\vec v_0 \cdot \vec x \, L[\vec x] \big)
\ee
Suppose as before that the Lagrangian is a function of $v^2 \equiv \dot{\vec x} \, ^2$ only,
\be
L[ \vec x \, ] = f(v^2) \; .
\ee
Then \eqref{Lorentz} reduces to the equation
\begin{align}
f' (v^2) \, 2 \vec v & \cdot \big[ \vec v_0  \big({v^2}/{c^2} -1\big) + \vec a \, (\vec v_0 \cdot \vec x)/c^2 \big] \\
& = f(v^2 ) \vec v_0 \cdot \vec v/c^2 + f'(v^2) 2 (\vec v \cdot \vec a)(\vec v_0 \cdot \vec x)/c^2 \; ,
\end{align}
where $\vec a \equiv \ddot {\vec x} $ is our particle's acceleration. The $\vec a$-dependent, $\vec x$-dependent terms cancel, and we are left with the differential equation
\be
2 f'(v^2) \big({v^2}/{c^2} -1\big) = f(v^2)/c^2 \; ,
\ee
whose general solution is
\be
f(v^2) \propto \sqrt{1- v^2/c^2}
\ee
The integration constant can be fixed by recalling that, by definition, at zero speed the Lagrangian must reduce to $-V_0 = -m c^2$. We thus find that the fundamental, invariant action must be
\be \label{free action}
S[\vec x \,] = -m c^2 \int dt \sqrt{1-\frac{\dot {\vec x} \, ^2}{c^2}} \; .
\ee

\vspace{0.5em}

\noindent\emph{\textbf{Interactions}} --- There's only so much one can say about free particles. They move in straight lines, at constant speeds. However, even before we consider interactions between different particles, it is worth noticing that the new putative space-time symmetry we have just unveiled depends explicitly on the value of $c^2 = V_0/m$. Then, for  one to be able to interpret such a symmetry as a geometric property of space-time that must be respected by the dynamics of all particles  inhabiting it, the value of $c^2$ must be the same for all particles. Notice also that, according to such a symmetry, the transformation law for time depends on $\vec x$. So, time will transform differently for different particles, but at least in a way that is completely determined by the positions they occupy. Although strange at first sight, this is not different from what happens for ordinary rotations in space: for rotations in the $xy$-plane, the $x$ coordinate of a particle transforms in a way that depends on its $y$ coordinate, and vice versa, and so the transformation laws of $x$ and $y$ are formally different for different particles, but  in a way that depends only on their positions.

We are now in a position to discuss interactions between different particles. Suppose that we have two particles with trajectories $\vec x_1(t)$ and $\vec x_2(t)$ and that we have measured that, at least at very small speeds, they interact through a distance-dependent potential. The low-speed interaction part of their action thus is
\be \label{Sint}
S_{\rm int} [\vec x_1, \vec x_2] \simeq -\int dt\,  V(r(t))   \; , 
\ee
with $r \equiv |\vec x_1 - \vec x_2|$. Such a term in the action is manifestly invariant under translations, rotations {\em and} the Galilei boosts \eqref{boost}, and we are usually quite happy about that. We shouldn't be. If the action is to be invariant under the new space-time symmetry we are postulating, under an active boost \eqref{boost} (or, more precisely, \eqref{new boost}), it should change by a total derivative, as in \eqref{Lorentz}. So, our new space-time symmetry predicts that \eqref{Sint} has to be corrected.

The interesting question is whether, like in the case of the free particle studied in the last section, the symmetry requirement is so powerful as to fully determine the invariant interaction action starting only from its zero speed limit, eq.~\eqref{Sint}. The answer is no. To see why, first we must recall that time transforms differently for  particles at different positions. And so, under our symmetry transformations, an instantaneous action at a distance like that described by \eqref{Sint} becomes a non-instantaneous action at a distance---that is, an interaction that depends on the two particles' positions evaluated at different times. In other words, action at a distance in space becomes action at a distance in space-time. This already tells us that for an interaction between two particles to be invariant
under our symmetry, it cannot be described by an action that is a single time integral of a Lagrangian that depends, through the dynamical variables, on a single time variable: for two particles, we need at least two time variables, $t_1$ and $t_2$, and if two pairs of spacetime points $(\vec x_1,t_1; \vec x_2, t_2)$ and $(\vec x \, '_1,t'_1; \vec x \, '_2, t'_2)$ can be related 
by a symmetry transformation, the interaction action
must treat them democratically. In particular, since our symmetry `rotates' time and space into each other, if the interaction depends on the spatial distance $|\vec x_1 - \vec x_2|$, it must also depend on the time distance $|t_1-t_2|$.
We are thus led to postulate that the interaction action takes the form
\be \label{bilocal}
S_{\rm int} [\vec x_1, \vec x_2] = \int dt_1 dt_2 \,  L_{\rm int}[\vec x_1, \vec x_2] (t_1, t_2) \; , 
\ee 
where $L_{\rm int}$ is a function of $\vec x_1(t_1)$ and $\vec x_2(t_2)$ and their derivatives (velocities, accelerations, and so on), whose specific form will be constrained by our symmetry. Crucially, it must be an explicit function of $t_1$ and $t_2$ themselves, more specifically, of their difference $t_1 - t_2$, because of the arguments above. In fact, as we will discuss in the next section, eq.~\eqref{bilocal} is not even the most general form the interaction action can take, but already at this level we see that the general problem of classifying the allowed interactions is quite complicated, and I will not attempt it here. 

What I will do instead, is to spell out the symmetry requirements on $L_{\rm int}$ assuming that the interaction does take the form \eqref{bilocal}, and then show that, starting from \eqref{Sint},  at very small speeds one can predict certain velocity-dependent corrections to potential interactions, but not all of the allowed ones. In order to do so, it is enough to generalize the argument that led us to \eqref{active} and \eqref{Lorentz} in the single time-integral case to our two-time integral case. The generalization is straightforward: we want that under the transformation
\begin{align}
 & t_a' = t_a - \frac{1}{c^2} \vec v_0 \cdot \vec x_a(t_a) \\
& \vec x  \, _a '(t_a') = \vec x_a(t_a) - \vec v_0 t_a \; ,
\end{align}
where $a=1,2$ labels the particles, the infinitesimal action element be invariant,
\be
dt_1' dt_2' L_{\rm int}[\vec x \,'_1 , \vec x \,' _2] (t'_1, t'_2) = dt_1 dt_2 L_{\rm int}[\vec x_1 , \vec x _2] (t_1, t_2) \; .
\ee
Following analogous manipulations to those in the previous section, this implies that under the active transformation
\be \label{xa'}
\vec x  \, _a '(t_a) = \vec x_a(t_a) - \vec v_0 t_a + \frac{1}{c^2} \dot {\vec x}_a(t_a)  \, \vec v_0 \cdot \vec x_a (t_a)
\ee
the interaction Lagrangian should change by
\begin{align} \label{symmetry}
\delta L_{\rm int}[\vec x_1 , \vec x _2]  = \frac{1}{c^2} \sum_a \frac{\partial}{\partial t_a} \big(\vec v_0 \cdot \vec x_a \,  L_{\rm int} [\vec x _1 , \vec x _2] \big)    \; ,
\end{align}
which is a direct generalization of \eqref{Lorentz}. Notice that, for notational simplicity, we are dropping the $(t_1, t_2)$ arguments, since now they are common to both sides of the equation. We must recall however that the interaction Lagrangian must depend on them not only through the dynamical variables, but also explicitly, because of the considerations above. 

Consider now the problem of extending \eqref{Sint} to small but non-zero speeds, accelerations, and so on. We can formally organize our interaction Lagrangian in \eqref{bilocal} as a power series in $1/c^2$:
\be
L_{\rm int} = L_{\rm int}^{(0)} + L_{\rm int}^{(1)} + L_{\rm int}^{(2)} + \dots \; , \qquad L_{\rm int}^{(n)} = {\cal O}(\sfrac{1}{c^2})^n \; ,
\ee
and impose our symmetry requirement \eqref{symmetry} order by order in $1/c^2$. Notice that, apart from the explicit factor of $1/c^2$ on the r.h.s~in \eqref{symmetry}, there is also a factor of $1/c^2$ in the active transformation of $\vec x_a(t_a)$ (eq.~\eqref{xa'}). So, at some fixed order $n$, eq.~\eqref{symmetry} reads
\begin{align} \label{n}
\delta_0 L^{(n)}_{\rm int} = -  \delta_1 L^{(n-1)}_{\rm int} +  \frac{1}{c^2} \sum_a \frac{\partial}{\partial t_a} \big(\vec v_0 \cdot \vec x_a \,  L^{(n-1)}_{\rm int}  \big)    \; ,
\end{align}
where $\delta_0$ and $\delta_1$ are, respectively, the zeroth order and  first order variations associated with \eqref{xa'},
and I am now dropping all arguments of $L_{\rm int}$ to minimize clutter.

Let's see what happens at the lowest orders.
Eq.~\eqref{Sint} gives us the zeroth order term of the interaction Lagrangian,
\be
L_{\rm int}^{(0)}[\vec x_1 , \vec x _2] (t_1, t_2) = - V\big(|\vec x_1(t_1) -  \vec x_2(t_2)|\big) \, \delta(t_1 - t_2) \; ,
\ee 
which, thanks to the delta function, is manifestly invariant under Galilei boosts \eqref{boost}, and so obeys our new symmetry requirement \eqref{n} to zeroth order, as expected. At first order, we must have
\begin{align} \label{n=1}
 \delta_0 L^{(1)}_{\rm int} &  = \delta_1\big( V(r) \big) \delta (t_1 -t_2) \\
  & - \frac{1}{c^2} \sum_a \frac{\partial}{\partial t_a} \big[\vec v_0 \cdot \vec x_a \,  V(r) \delta (t_1 -t_2)  \big]  \nonumber \\
  & = -  \frac{1}{c^2} \vec v_0 \cdot (\vec v_1+ \vec v_2) \,  V(r) \delta (t_1 -t_2) \nonumber \\
  & -  \frac{1}{c^2}  V(r)  \vec v_0 \cdot  \sum_a \vec x_a \, \frac{\partial}{\partial t_a}  \delta (t_1 -t_2) \nonumber  \; ,
\end{align}
where $\vec x_a$ and $\vec v_a \equiv \dot {\vec x}_a$ are evaluated at their own $t_a$, and $\vec r \equiv \vec x_1(t_1) - \vec x_2(t_2)$.
Notice that there was a perfect cancellation between the $\delta_1$ variation of $V$ and the terms in which the time-derivatives act on $V$. This is because the ${\cal O}(1/c^2)$ term in the transformation law \eqref{xa'} secretly comes from a transformation of $t_a$. At higher orders, for interactions involving velocities and higher derivatives of the $\vec x_a$'s, such a cancellation is only partial.

To proceed further, it is convenient to introduce for our two particles a common time variable and a relative one,
\begin{align} \label{time}
& t \equiv \sfrac12(t_1+t_2) \; , \quad \Delta t \equiv t_1 - t_2 \quad \Rightarrow \quad
t_{1,2} = t\pm \sfrac12 \Delta t \; .
\end{align}
Then, the last line of \eqref{n=1} becomes
\be
-  \frac{1}{c^2} \vec v_0 \cdot \vec r \,  V(r)   \delta' (\Delta t) 
\ee
Using the distributional identity\footnote{Despite appearances, this identity does not secretely involve integrating by parts. In fact, it holds for all test functions that are differentiable at $y=0$, regardless of whether they decay at infinity. To see this, it is enough to take the derivative of the more familiar identity $f(y) \delta(y) = f(0) \delta(y)$.}
\begin{align} \label{identity}
f(y) \delta'(y) = f(0) \delta' (y) - f'(y) \delta (y) \; ,
\end{align}
where $f$ is any differentiable function, and 
\be
\frac{\partial}{\partial \Delta t} \vec r = \frac{1}{2} \big(\vec v_1 + \vec v _2 \big) \equiv \vec v_{\rm avg}
\ee
(`avg' for `average'),
this can be rewritten as
\begin{align}
& -  \frac{1}{c^2} \big[ \vec v_0 \cdot \vec r \,  V(r) \big]_{\Delta t=0} \,  \delta' (\Delta t) \\
& + \frac{1}{c^2}  v_0^ i   v_{\rm avg}^j  \big[\delta^{ij} \,  V(r) + \hat r^i \hat r^j  \, r V'(r)\big] \,  \delta (\Delta t) \nonumber \; .
\end{align}

Putting everything together, we get that  under Galilei boosts the first-order interaction Lagrangian must change by
\begin{align} \label{n=1 final}
 \delta_0 L^{(1)}_{\rm int} &  =- \frac{1}{c^2}  v_0^ i   v_{\rm avg}^j  \big[\delta^{ij} \,  V(r) - \hat r^i \hat r^j  \, r V'(r)\big] \,  \delta (\Delta t)   \nonumber \\
  & -  \frac{1}{c^2} \big[ \vec v_0 \cdot \vec r \,  V(r) \big]_{\Delta t=0} \,  \delta' (\Delta t)  \; .
\end{align}
We can now reconstruct the most general $L^{(1)}_{\rm int} $ compatible with this symmetry requirement. Because of translational invariance and rotational invariance, $L^{(1)}_{\rm int} $ can depend on the trajectories $\vec x_a(t_a)$ through rotationally invariant combinations of $\vec r$, $\vec v_a$, $\vec a_a \equiv \ddot {\vec x}_a$, and higher time derivatives of $\vec x_a$ (collectively denoted below by `\dots'). Parametrizing the explicit time-dependence of $L^{(1)}_{\rm int} $ in terms of $\delta (\Delta t)$ and its derivatives, and using the identity \eqref{identity} and its higher-order generalizations, all these quantities can be evaluated at $\Delta t = 0$. Furthermore, without loss of generality, we can trade in the dependence on the individual $\vec v _a$'s for that on $\vec v_{\rm avg}$ and $\Delta \vec v \equiv v_1 - \vec v _2$. Then, all the quantities $L^{(1)}_{\rm int} $  can depend on are invariant under Galilei boosts, apart from  $\vec v_{\rm avg}$, which shifts by
\be
\vec v_{\rm avg} \to \vec v_{\rm avg} - \vec v_0 \; .
\ee
So, for \eqref{n=1 final} to be obeyed, the $L^{(1)}_{\rm int} $ we are after must take the form
\begin{align}
L^{(1)}_{\rm int} & (\vec r, \vec v_{\rm avg}, \Delta \vec v, \vec a_a, \dots; \Delta t)  =  \\ 
& +\frac{1}{2c^2}\big[  \big( v^2_{\rm avg} \,  V(r) - (\vec v_{\rm avg} \cdot \hat r)^2  \, r V'(r) \big) \delta (\Delta t) \nonumber \\
& + f(\vec r,  \Delta \vec v, \vec a_a, \dots) \delta (\Delta t)   \nonumber \\
& +  2 \big( \vec v_{\rm avg} \cdot \vec r \,  V(r) \big)_{\Delta t=0} \,  \delta' (\Delta t) \nonumber  \\
& +  \sum_{n \ge1} g_n(\vec r,  \Delta \vec v, \vec a_a, \dots)_{\Delta t=0} \, \delta^{(n)} (\Delta t) \big] \nonumber \; ,
\end {align}
for arbitrary $f$ and $g_n$.

So, the symmetry requirement \eqref{n=1 final} completely determines the dependence on $\vec v_{\rm avg}$, but leaves everything else arbitrary. Notice that the action \eqref{bilocal} involves an integral over $dt_1 dt_2$ of the Lagrangian, or, equivalently, over $dt d \Delta t$. After performing the integral in $\Delta t$, all terms involving derivatives of $\delta(\Delta t)$ disappear, and we thus reach the conclusion that, at ${\cal O}(1/c^2)$, a potential interaction of the form  \eqref{Sint} must be corrected as
\begin{align} 
S_{\rm int} & [\vec x_1, \vec x_2] \simeq \int dt \Big\{ -V(r) \nonumber \\
& +\frac1{2c^2} \big[ v^2_{\rm avg} \,  V(r) - (\vec v_{\rm avg} \cdot \hat r)^2  \, r V'(r)   \label{corrected action} \\
&  + f(\vec r,  \Delta \vec v, \vec a_a, \dots) \big]\Big\} \; .   \nonumber
\end{align} 

In fact, on physical grounds, one expects the power expansion in $1/c^2$ to be related to a power expansion in terms of velocities, accelerations, and so on, that is, in terms of time derivatives. Then, on dimensional grounds, one expects $f$ to contain terms of second order in $\Delta v$ and of first order in the $a_a$'s, and no higher time derivatives,
\be
f \sim \Delta v^2 V(r) +   a_a r \,  V(r) \; .
\ee

\vspace{0.5em}

\noindent\emph{\textbf{Back to the future}} --- 
We can now stop pretending that we don't know about special relativity. Our results for free particles---say, the action \eqref{free action}---are of course a rewriting of standard Lorentz invariance. Still, I find it amusing that they are also consistent with a Euclidean $SO(4)$ spacetime symmetry (for negative $c^2$), which would treat completely democratically all spacetime coordinates, without giving time any special status. Had special relativity been discovered, or postulated, in the way presented here, one could have probably ruled out the $SO(4)$ possibility by causality arguments. For instance, in the $SO(4)$ case, by a finite space-time rotation of angle $v_0/c = \pi/2$, one can turn a static trajectory ($\vec x = {\rm const}$) into an instantaneous one ($t = {\rm const}$); for larger rotation angles, one can flip the time-orientation of the trajectory.

Less trivial, and potentially useful, are our results about interactions. Usually we say that special relativity is inconsistent with action at a distance, and we introduce local fields in order to mediate long-range interactions between particles in a way consistent with Lorentz invariance. That, as far as we know, is the physically correct fundamental description of long-range interactions. Still, if we integrate out these local fields, we must end up with a non-local action for our particles, and this must be consistent with Lorentz invariance. In the case of a bi-local interaction action of the form \eqref{bilocal}, eq.~\eqref{symmetry} is the symmetry requirement that must be satisfied.

Notice that for fields with self-interactions, such as the gravitational one, the interaction action mediated by such fields will be more general than \eqref{bilocal}. In perturbation theory, it will be the sum of a bi-local one, a tri-local one, and so on. The reason is that, treating our particles as external sources for the field mediating interactions, there will be connected Feynman diagrams in which the field attaches several times to our particles' trajectories and interact with itself through some local vertices. This phenomenon goes hand in hand with our field's being able to mediate three-body interactions, four-body interactions, and so on---the same Feynman diagrams we just described can be used to connect more than two trajectories.  So, we can combine these two physically distinct cases into an $N$-body generalization of \eqref{bilocal},
\begin{align} \label{N-local}
S_{\rm int}[\vec x_1, & \dots, \vec x_N]   =\\
& \int dt_1 \dots dt_N \,  L_{\rm int}[\vec x_1, \dots, \vec x_N] (t_1, \dots,  t_N) \; ,  \nonumber 
\end{align}
and of \eqref{symmetry}, 
\be \label{N-symmetry}
\delta L_{\rm int}[\vec x_1 , \dots \; , \vec x _N]  = \frac{1}{c^2} \sum_a \frac{\partial}{\partial t_a} \big(\vec v_0 \cdot \vec x_a \,  L_{\rm int} [\vec x _1 , \dots , \vec x _N] \big)    \; ,
\ee
with the understanding that some of the trajectories $\vec x_a(t_a)$ could in fact be the same one, but evaluated at different times.
The statement then is:  if at some order in perturbation theory our field mediates up to an $N$-body potential in the non-relativistic limit, then at the same in order in perturbation theory the relativistic generalization of the potential interactions mediated by such a field must take the form \eqref{N-local}, and must obey \eqref{N-symmetry}.

However, such complications appear to be irrelevant in a time-derivative expansion for nearly static trajectories. The reason is that, as clear from inspecting the first example we  present below, to any finite order in a time-derivative expansion the generalized interaction Lagrangian \eqref{N-local} will explicitly depend on the time differences $t_1-t_2$, \dots, $t_1-t_N$ through delta-functions and derivatives thereof. In the case of two trajectories, this allows one to perform the $t_3, \dots, t_N$ integrals explicitly, upon which one is left with a bilocal interaction Lagrangian of the form \eqref{bilocal}, which must obey our symmetry requirement \eqref{symmetry} (to the correct order in $1/c^2$). In particular, to first order in $1/c^2$, after integrating in the time difference $t_1-t_2$, one can still expect the interaction action to take the form \eqref{corrected action} \footnote{This simple argument neglects the complications associated with genuinely nonlocal-in-time ``tail" effects, which appear in certain high-order computations---see e.g.~\cite{DJS,FPRS}. Such effects are associated with non-analyticities in the low-frequency expansion, due to the on-shell emission of the massless field mediating the long-distance interactions we are after. Those effects will need to be studied separately from the analysis presented here.}.

We can now run a number of checks on our first order result, eq.~\eqref{corrected action}. The claim is that, in a Lorentz-invariant theory, at small but nonzero velocities, interactions between point particles must take that form regardless of the nature (spin, self-interactions) of the field mediating them. From now on, I will set $c$ to one.

Let's start with a long-range interaction mediated by a scalar local operator ${\cal O}(x)$. Let's assume that ${\cal O}$ has local monopole-type couplings to our trajectories,
\be
S \supset g_1 \int d \tau_1 \, {\cal O}(x_1) +  g_2 \int d \tau_2 \, {\cal O}(x_2) \; .
\ee 
Such an operator could be a free or weakly coupled fundamental scalar field, a scalar field with strong self-interactions, or even a composite operator. I will make no assumptions in this sense. I will only assume, for simplicity, that $g_1$ and $g_2$ are small (in the appropriate units), so that we can work to first non-trivial order in them. Then, at order $g_1 g_2$, the effective interaction mediated by ${\cal O}$ takes the form
\be
S_{\rm int}[\vec x_1, \vec x_2] = \int \frac{d^4 k}{(2\pi)^4} \tilde J_1(-k) \, \tilde G(k^2) \, \tilde J_2(k) \; ,
\ee
where $G$ is ($i$ times) ${\cal O}$'s two-point function\footnote{For our derivation it does not matter which two-point function (Feynman, retarded, etc.) we choose, as long as $G$ is a Green's function for ${\cal O}$: $\langle {\cal O} \rangle_J = G * J$.}, and $J_a$ is the source for ${\cal O}$ associated with particle $a$, as determined by the local couplings above:
\be
J_a(x) =  g_a \int d \tau_a \delta^4(x - x_a(\tau_a)) \; ,
\ee
or, equivalently,
\be
\tilde J_a(k) =  g_a \int d \tau_a  e^{-i k \cdot x_a(\tau_a)} \; .
\ee
We thus get
\be
S_{\rm int}[\vec x_1, \vec x_2] = g_1 g_2 \int d \tau_1d \tau_2\frac{d^4 k}{(2\pi)^4}  \tilde G(k^2) e^{i k\cdot (x_1(\tau_1) - x_2(\tau_2)) } \; .
\ee
This is a fully relativistic expression, and manifestly so. For very low speeds, it is convenient to rewrite the integrals in $\tau_{1,2}$ as integrals in time,
\be
d\tau_{a} = \sqrt{1- v_a^2} \, dt_a \; ,
\ee 
and to expand the two-point function of ${\cal O}$ in powers of  frequency:
\be
\tilde G(\vec k \, ^2 -\omega^2) = \tilde G(\vec k \, ^2 ) - \tilde G'(\vec k \, ^2) \omega^2 + \dots \; .
\ee
Then, to order $v^2$ or $\omega^2$, we get
\begin{align}
S_{\rm int}& [\vec x_1, \vec x_2] \simeq g_1 g_2 \int d t_1 d t_2 \frac{d^4 k}{(2\pi)^4} \Big[ \tilde G(\vec k \, ^2) \nonumber \\
& - \sfrac12 \tilde G(\vec k \, ^2 )  (v_1^2(t_1) + v_2^ 2(t_2)) - \tilde G'(\vec k \, ^2) \omega^2 \Big] \\
& \times e^{-i \omega (t_1 - t_2) }e^{i \vec k\cdot (\vec x_1(t_1) - \vec x_2(t_2)) } \nonumber
\; .
\end{align}
The integral in $\omega$ can be performed explicitly, yielding a $\delta(t_1 -t _2)$ term and a $\delta''(t_1-t_2)$ one. Rewriting the latter  as $\delta''(t_1-t_2) = -\partial_{t_1} \partial_{t_2} \delta(t_1-t_2)$, performing the integral in $t_2$, and relabeling $t_1 \to t$, we are left with
\begin{align} \label{expanded O}
S_{\rm int}& [\vec x_1, \vec x_2] \simeq g_1 g_2 \int d t  \frac{d^3 k}{(2\pi)^3} \Big[ \tilde G(\vec k \, ^2) \\
& - \sfrac12 \tilde G(\vec k \, ^2 )  (v_1^2+ v_2^ 2) - v^i_1 v^j_2 \,  k^i  k^j  \tilde G'(\vec k \, ^2)  \Big]  e^{i \vec k \cdot \vec r }  \nonumber 
\; ,
\end{align}
where $\vec v_1$, $\vec v_2$, and $\vec r \equiv x_1 - \vec x_2$ are all evaluated at $t$. 

For vanishing velocities, such an interaction action defines the static potential mediated by ${\cal O}$. That is, 
\be
V(r) \equiv - g_1 g_2 \int  \frac{d^3 k}{(2\pi)^3} \,  \tilde G(\vec k \, ^2)   e^{i \vec k \cdot \vec r } \; .
\ee
Then, it's clear that the first velocity-dependent correction in \eqref{expanded O} is also proportional to the potential. For the second, we have to manipulate it a bit: using
\be
\vec \nabla _{\vec k} \tilde G(\vec k \, ^2) = 2 \vec k \, \tilde G'(\vec k \, ^2) \; , 
\ee
we have
\begin{align}
g_1 g_2 \int  \frac{d^3 k}{(2\pi)^3} & \,  k^i  k^j   \tilde G'(\vec k \, ^2) e^{i \vec k \cdot \vec r }  \\
& = \sfrac12 g_1 g_2 \int  \frac{d^3 k}{(2\pi)^3} \,  k^i \nabla^j_{\vec k}  \tilde G(\vec k \, ^2) e^{i \vec k \cdot \vec r } \nonumber \\
& =  \sfrac12 \nabla^i \big( r^j V(r)\big) \nonumber \\
& =  \sfrac12 \big( \delta^{ij} V(r) + \hat r^i \hat r^j \, r V'(r)\big) \; . \nonumber
\end{align}

Then, putting everything together, we have
\begin{align} \label{expanded O}
S_{\rm int}& [\vec x_1, \vec x_2] \simeq  \int d t   \Big[ -V(r) + \sfrac12 V(r)  (v_1^2+ v_2^ 2)  \nonumber \\
&  - \sfrac12 \big( \delta^{ij} V(r) + \hat r^i \hat r^j \, r V'(r)\big) v^i_1 v^j_2  \Big]     
\; ,
\end{align}
Using
\be \label{v1,2}
\vec v_{1,2} = \vec v_{\rm avg} \pm \sfrac12 \Delta \vec v \; ,
\ee
we finally get 
\begin{align}
S_{\rm int}& [\vec x_1, \vec x_2] \simeq  \int d t   \Big[ -V(r)  \nonumber \\ 
& + \sfrac12 \big( V(r) \delta^{ij}- \hat r^i \hat r^j \, r V'(r)\big) v_{\rm avg} ^i v_{\rm avg}^j \label{spin0} \\
&  + \sfrac18 \big( 3 \, \delta^{ij} V(r) + \hat r^i \hat r^j \, r V'(r)\big) \Delta v^i  \Delta v^j  \Big]    \nonumber 
\; ,
\end{align}
which is precisely of the predicted form, eq.~\eqref{corrected action}. We stress again that this result holds for monopole-monopole interactions mediated by any scalar operator ${\cal O}(x)$---fundamental or composite, weakly coupled or not.

Moving up in the spin of the field that mediates our long-distance interactions, we can now turn our attention to electromagnetism. For two point-particles of charges $q_1, q_2$, integrating out the electromagnetic field  to order $v^2/c^2$ yields the so-called Darwin Lagrangian (see e.g.~\cite{Jackson}),
\be \label{Darwin}
S_{\rm int} [\vec x_1, \vec x_2] \simeq  \int d t \, V(r) \big\{-1 + \sfrac12\big[ \vec v_1 \cdot \vec v_2 + (\vec v_1 \cdot \hat r ) (\vec v_2 \cdot \hat r )\big]\big\}  
\ee
where
\be
V(r)  \equiv \frac{q_1 q_2}{r} 
\ee
is the electrostatic potential energy. Using \eqref{v1,2}, we get
\begin{align}
S_{\rm int} & [\vec x_1, \vec x_2]  \simeq  \int d t \, \big\{-V(r)  \nonumber \\
& + \sfrac12 V(r)\big[  v_{\rm avg} ^{\, 2} + (\vec v_{\rm avg} \cdot \hat r )^2 \big] \label{spin1} \\
& - \sfrac18 V(r)\big[ \Delta v ^{\, 2} + (\Delta \vec v \cdot \hat r)^2 \big] \big\}  \; , \nonumber
\end{align} 
which is again of the predicted form, eq.~\eqref{corrected action}, since now $V(r) \propto 1/r$.

Next, we can consider the interactions of two point-masses in general relativity (GR). The problem of systematically expanding GR
at low speeds compared to the speed of light has a long history, and, in typical situations, turns out be tied to an expansion in powers of $G$. This combined expansion, with ${\cal O}(v^2/c^2) = {\cal O}(G)$,  goes under the name of Post-Newtonian (PN) expansion, and has been recast more recently in effective field theory terms with an eye towards applications to gravitational wave physics \cite{GR}. Our prediction \eqref{corrected action} should be valid to order $v^2/c^2$, but up to arbitrary order in $G$ (as long there is an instantaneous potential $V(r)$). So, given a result at $n$-th PN order (or, in short, ``at $n$PN"), with $n=0$ corresponding to the Newtonian limit, we can check whether its ${\cal O}(v^2 G^n)$ terms agree with our eq.~\eqref{corrected action} when $V(r)$ is taken from the ${\cal O}(v^0 G^n)$ terms at $(n-1)$PN. 

At 1PN, the interactions between two point-masses $m_1$ and $m_2$ are given by the Einstein-Infeld-Hoffman action \cite{EIH} (see also \cite{GR}),
\begin{align} \label{EIH}
S^{\rm 1PN}_{\rm int} & [\vec x_1, \vec x_2]  \supset -  \int d t \, V^{\rm 0PN}(r)  \\
&  \times \sfrac12 \big[  3 (v_1^2 + v_2^2) -  7 \vec v_1 \cdot \vec v_2 - (\vec v_1 \cdot \hat r ) (\vec v_2 \cdot \hat r ) \big] \; , \nonumber
\end{align} 
where 
\be \label{V0PN}
V^{\rm 0PN}(r) \equiv- G \frac{m_1 m_2}{r} 
\ee
is just the Newtonian potential energy. Using our parametrization \eqref{v1,2}, we get
\begin{align}
S^{\rm 1PN}_{\rm int} & [\vec x_1, \vec x_2]  \supset \nonumber \\ 
&  \int d t \, \big \{ \sfrac12 V^{\rm 0PN}(r)  \big[  v_{\rm avg} ^{\, 2} + (\vec v_{\rm avg} \cdot \hat r )^2 \big] \label{spin2} \\
& \quad - \sfrac18 V^{\rm 0PN}(r) \big[ 13 \Delta v ^{\, 2} + (\Delta \vec v \cdot \hat r)^2 \big] \big\} \; , \nonumber
\end{align} 
which is of the correct form \eqref{corrected action}, since $V^{\rm 0PN}(r) \propto 1/r$.

At 2PN, the interaction action has been computed in \cite{GilmoreRoss}. Its ${\cal O}(v^2 G^2)$ terms are
\begin{align}
S^{\rm 2PN}_{\rm int} & [\vec x_1, \vec x_2]  \supset  \int d t \, \big\{V^{\rm 1PN}_{12}(r)  \big[ 7 (\vec v_1 \cdot \hat r )^2 + (\vec v_2 \cdot \hat r )^2    \nonumber \\
& -7 (\vec v_1 \cdot \hat r ) (\vec v_2 \cdot \hat r )
+\sfrac12 v_1^2 + \sfrac72 v_2^2 -\sfrac72 \vec v_1 \cdot \vec v_2  \big]  \label{2PN}\\
& + (1 \leftrightarrow 2) \; , \nonumber
\end{align} 
where
\be
V^{\rm 1PN}_{ab}(r) \equiv G^2 \frac{m_a^2 m_b}{2r^2} \; , \qquad a,b=1,2 \; ,
\ee
and the full 1PN potential energy is
\be
V^{\rm 1PN}(r) = V^{\rm 1PN}_{12}(r) + V^{\rm 1PN}_{21}(r) \; .
\ee
In terms of $\vec v_{\rm avg}$ and $\Delta \vec v$, we get
\begin{align}
& S^{\rm 2PN}_{\rm int}  [\vec x_1, \vec x_2]  \supset \nonumber  \\ 
&  \int d t \, \big \{ \sfrac12 V^{\rm 1PN}(r)  \big[  v_{\rm avg} ^{\, 2} + 2 (\vec v_{\rm avg} \cdot \hat r )^2 \big] \label{spin2-2PN}\\
& \quad\quad + \sfrac{15}{8} V^{\rm 1PN}(r) \big[  \Delta v ^{\, 2} +2  (\Delta \vec v \cdot \hat r)^2 \big] \nonumber  \\
& -3 \, \Delta V^{\rm 1PN}(r) \,  \big[\vec v_{\rm avg} \cdot \Delta \vec v - 2 ( \vec v_{\rm avg} \cdot \hat r) (\Delta \vec v \cdot \hat r)\big]\big\}\; , \nonumber
\end{align} 
where
\be
\Delta V^{\rm 1PN}(r) \equiv V^{\rm 1PN}_{12}(r)   - V^{\rm 1PN}_{21}(r)\; .
\ee
The first two lines of the r.h.s~match our \eqref{corrected action}, since now $V^{\rm 1PN}(r) \propto 1/r^2$, but the third line involves mixed $v_{\rm avg} \Delta v$ terms, which are not contemplated by \eqref{corrected action}.
However, upon integrating by parts, that line can be rewritten as
\be
3 \, \Delta V^{\rm 1PN}(r) \, \vec a_{\rm avg} \cdot \vec r  \; ,
\ee
where $\vec a_{\rm avg} \equiv \sfrac12 (\vec a_1 + \vec a_2) $ is the average acceleration. Such an expression {\em is} compatible with eq.~\eqref{corrected action}, which tells us that a more ``Lorentz-friendly" version of  \eqref{2PN} is
\begin{align}
S^{\rm 2PN}_{\rm int} & [\vec x_1, \vec x_2]  \supset  \int d t \, \big\{V^{\rm 1PN}(r)  \big[ 2 (v_1^2+v_2^2) \nonumber \\ 
& + 4 \big( (\vec v_1 \cdot \hat r)^2 +(\vec v_2 \cdot \hat r )^2 \big) -\sfrac72 \vec v_1 \cdot \vec v_2 \\
&  -7 (\vec v_1 \cdot \hat r ) (\vec v_2 \cdot \hat r )  \big]  \nonumber \\
& + \sfrac32 \Delta V^{\rm 1PN}(r) \, \big(\vec a_1 \cdot \vec r + \vec a_2 \cdot \vec r \,  \big) \big\} \; . \nonumber
\end{align} 

We could keep going---the interaction action between two point-masses in GR  has  been computed up to 4PN \cite{FPRS}---but, for the sake of brevity, I will stop here.

As a final example, we can consider scalar-tensor theories of gravity, where a PN expansion can also be carried out. At 1PN, the ${\cal O}(v^2 G)$ terms read \cite{KPV, DEF}
\begin{align} 
S^{\rm 1PN}_{\rm int} & [\vec x_1, \vec x_2]  \supset -  \int d t \, V^{\rm 0PN}(r)  
 \sfrac12 \big[ (1+ 2 \gamma) (v_1^2 + v_2^2) \nonumber \\
 & -  (3 + 4 \gamma) \vec v_1 \cdot \vec v_2 - (\vec v_1 \cdot \hat r ) (\vec v_2 \cdot \hat r )   \big] \; , \label{ST}
\end{align} 
where the potential $V^{\rm 0PN}$ still takes the form \eqref{V0PN} (with an effective $G$), and $\gamma$ is a free parameter. At this order, however, such terms are nothing but a linear combination of the corresponding ones in the scalar case, eq.~\eqref{expanded O}, and in the GR case, eq.~\eqref{EIH}, with coefficients $\frac12(1\mp \gamma)$. Then, since those cases obey our symmetry constraint, so does eq.~\eqref{ST}.

\vspace{1em}

\noindent\emph{\textbf{Concluding remarks}} ---
The first part of the paper is clearly a historical fantasy, perhaps of limited interest and value. 
However, it motivates a possibly useful, novel viewpoint on long-range Lorentz-invariant interactions between point particles: yes, we know that fundamentally these are mediated by local fields; but we can still work out systematically the consequences of Lorentz invariance for the non-local interactions we are left with after these fields have been integrated out. Moreover, we can do so by only using the transformation properties of kinematical variables under {\em Galilean} boosts, which are extremely simple.

In a $1/c^2$ expansion, the recursive symmetry requirement takes the general form \eqref{n}.
For two particles with monopole-monopole potential interactions, the first velocity-dependent corrections to their interactions must take the form \eqref{corrected action}. In fact, parameterizing the interaction action using the same variables as spelled out after eq.~\eqref{n=1 final}, one can see from \eqref{n} that at {\em any} order in the $1/c^2$ expansion the dependence on the average velocity $\vec v_{\rm avg}$ is completely determined by the previous orders. The fundamental reason is that \eqref{n} states  how the $n$-th order action term should transform under Galilean boosts. One can then organize the kinematical variables entering the action in such a way that $\vec v_{\rm avg}$ is the only one transforming non-trivially under Galilean boosts.

Interestingly, Lorentz invariance fixes directly the dependence on $\vec v_{\rm avg}$ rather than on the center-of-mass (CM) velocity, which usually has a more prominent status. For instance, for unequal masses, the CM velocity is conserved in the absence of external perturbations, but the average velocity is not. What's perhaps good for us about this is that, in an expansion in $1/c^2$, the average velocity is fixed once and for all, while the CM velocity is not. The reason is that, in a relativistic theory, the CM velocity is
\be
v^i_{\rm CM} = \frac{P^i}{P^0}= \frac{\int d^3 x \, T^{0i}}{\int d^3 x \, T^{00}} \; , 
\ee
and so each new correction to $P^\mu$ (or $T^{\mu\nu}$) corrects the CM velocity as well.
Using the average velocity rather than the CM one makes a recursive application of \eqref{n} straightforward.

Notice that in the spin-0, spin-1, and spin-2 examples of last section, the interaction actions~\eqref{spin0}, \eqref{spin1}, and \eqref{spin2}  all exhibit the same coefficient for one of the $\Delta \vec v$-dependent terms,
\be
S_{\rm int} \supset \int dt \,  \sfrac18 r V'(r) \,  (\Delta \vec v \cdot \hat r)^2 \; .
\ee
This is surprising, because such a term is unconstrained by Lorentz invariance, and so one expects different fields to yield different coefficients for it. (That is the case for instance for the other $\Delta \vec v$-dependent term, $\Delta v^2$.) In fact, this is an accident of the lowest order in perturbation theory for the couplings between our particles and the mediator field, as one can check by inspecting the 2PN result \eqref{spin2-2PN}, where that term has a different coefficient. 
In our computations, the accident stems from the fact that the only $(\vec v_a \cdot \hat r)(\vec v_b \cdot \hat r)$ terms in the original actions \eqref{expanded O}, \eqref{Darwin}, \eqref{EIH} are of the mixed $v_1v_2$ form, and never of the form $v_1 v_1$ or $v_2 v_2$, thus tying the coefficient of $(\Delta \vec v \cdot \hat r)^2$ to that of $(\vec v_{\rm avg} \cdot \hat r)^2$, which is completely determined by Lorentz invariance.
Why only the mixed form for these terms is allowed to lowest order in perturbation theory is not immediately obvious to me, but it can probably be understood diagramatically, taking into account the remarks around eqs.~\eqref{N-local} and \eqref{N-symmetry}.

I hope the approach and results presented here---perhaps suitably extended to include the effects of spin---will be useful in bootstrapping, or at least checking, some of the computations needed for the post-Newtonian expansion of general relativity. In that context, whether a final result is compatible with Lorentz-invariance is usually checked using the transformation laws derived in \cite{BF}. Those transformation laws, however, are quite complicated, which makes using them at high orders intimidating. (This is perhaps the reason why the check of Lorentz invariance is usually alluded to but not reported explicitly in the recent literature on the PN expansion.) In contrast, as emphasized above, the recursive formula \eqref{n} only uses the transformation properties of the kinematical variables under Galilei boosts, which, in the right variables, are almost trivial. 

As an example, consider the 5PN static potential derived in \cite{FS},
\be \label{static 5PN}
V^{\rm 5PN}(r) \propto \frac{G^6}{r^6}  \; ,
\ee
where the omitted proportionality constant is a known seventh order polynomial of the two masses.
Our results tell us immediately that, still at 5PN, this must be corrected by ${\cal O}(v^2 G^5)$ terms of the form
\begin{align}
S^{\rm 5PN} & \supset \int dt \, \sfrac1{2} \big[ v^2_{\rm avg} \,  V^{\rm 4PN}(r) +5 (\vec v_{\rm avg} \cdot \hat r)^2  \, V^{\rm 4PN}(r)  \nonumber \\
&  + f^{\rm 5PN}(\vec r,  \Delta \vec v, \vec a_a, \dots) \big] \; ,   \label{5PN} 
\end{align}
where $V^{\rm 4PN}(r)$ is the 4PN static potential \cite{FPRS},
\be
V^{\rm 4PN}(r) \propto \frac{G^5}{r^5}
\ee
(omitting again a known mass-dependent proportionality constant), and $ f^{\rm 5PN}$ is generically of order
\be
f^{\rm 5PN} \sim \Delta v^2 V^{\rm 4PN}(r) +   a_a r \,  V^{\rm 4PN}(r) \; .
\ee
Moreover, we can use \eqref{static 5PN} to predict the form of the ${\cal O}(v^2 G^6)$ terms at 6PN:
\begin{align}
S^{\rm 6PN} & \supset \int dt \, \sfrac1{2} \big[ v^2_{\rm avg} \,  V^{\rm 5PN}(r) +6 (\vec v_{\rm avg} \cdot \hat r)^2  \, V^{\rm 5PN}(r)  \nonumber \\
&  + f^{\rm 6PN}(\vec r,  \Delta \vec v, \vec a_a, \dots) \big] \; ,   \label{6PN} 
\end{align}
with $f^{\rm 6PN}$  of order
\be
f^{\rm 6PN} \sim \Delta v^2 V^{\rm 5PN}(r) +   a_a r \,  V^{\rm 5PN}(r)  \; .
\ee

Although we stopped at first order in $1/c^2$, using recursively eq.~\eqref{n} one can determine completely the dependence on $\vec v_{\rm avg}$ to any given order in terms of lower order action terms. On the other hand, the dependence on the other kinematical variables seems to be unconstrained by Lorentz invariance.

\vspace{1em}

\begin{acknowledgments}
\noindent\textbf{\emph{Acknowledgments}} --- 
I am grateful to Adrien Kuntz, Federico Piazza, and Riccardo Rattazzi for useful discussions and comments. This work is partially supported by the US DOE (award number DE-SC011941) and by the Simons Foundation (award number 658906).
\end{acknowledgments}

\bibliographystyle{apsrev4-1}

\begin{thebibliography}{99}

\bibitem{borges}
Jorge Luis Borges,
``Pierre Menard, Autor del Quijote'',
Sur 56 (May 1939): 7--16.

\bibitem{DJS}
T.~Damour, P.~Jaranowski and G.~Sch\"afer,
``Nonlocal-in-time action for the fourth post-Newtonian conservative dynamics of two-body systems,''
Phys. Rev. D \textbf{89}, no.6, 064058 (2014)
doi:10.1103/PhysRevD.89.064058
[arXiv:1401.4548 [gr-qc]].

\bibitem{FPRS}
S.~Foffa, R.~A.~Porto, I.~Rothstein and R.~Sturani,
``Conservative dynamics of binary systems to fourth Post-Newtonian order in the EFT approach II: Renormalized Lagrangian,''
Phys. Rev. D \textbf{100}, no.2, 024048 (2019)
doi:10.1103/PhysRevD.100.024048
[arXiv:1903.05118 [gr-qc]].


\bibitem{Jackson} 
  J.~D.~Jackson,
  ``Classical Electrodynamics,''
  Wiley, 1998.

\bibitem{GR} 
  W.~D.~Goldberger and I.~Z.~Rothstein,
  ``An Effective field theory of gravity for extended objects,''
  Phys.\ Rev.\ D {\bf 73}, 104029 (2006)
  doi:10.1103/PhysRevD.73.104029
  [hep-th/0409156].

\bibitem{EIH} 
  A.~Einstein, L.~Infeld and B.~Hoffmann,
  ``The Gravitational equations and the problem of motion,''
  Annals Math.\  {\bf 39}, 65 (1938).
  doi:10.2307/1968714

\bibitem{GilmoreRoss} 
  J.~B.~Gilmore and A.~Ross,
  ``Effective field theory calculation of second post-Newtonian binary dynamics,''
  Phys.\ Rev.\ D {\bf 78}, 124021 (2008)
  doi:10.1103/PhysRevD.78.124021
  [arXiv:0810.1328 [gr-qc]].

\bibitem{KPV}
A.~Kuntz, F.~Piazza and F.~Vernizzi,
``Effective field theory for gravitational radiation in scalar-tensor gravity,''
JCAP \textbf{05}, 052 (2019)
doi:10.1088/1475-7516/2019/05/052
[arXiv:1902.04941 [gr-qc]].

\bibitem{DEF}
T.~Damour and G.~Esposito-Farese,
``Tensor multiscalar theories of gravitation,''
Class. Quant. Grav. \textbf{9}, 2093-2176 (1992)
doi:10.1088/0264-9381/9/9/015

\bibitem{BF}
L.~Blanchet and G.~Faye,
``Lorentzian regularization and the problem of point - like particles in general relativity,''
J. Math. Phys. \textbf{42}, 4391-4418 (2001)
doi:10.1063/1.1384864
[arXiv:gr-qc/0006100 [gr-qc]].

\bibitem{FS}
S.~Foffa, P.~Mastrolia, R.~Sturani, C.~Sturm and W.~J.~Torres Bobadilla,
``Static two-body potential at fifth post-Newtonian order,''
Phys. Rev. Lett. \textbf{122}, no.24, 241605 (2019)
doi:10.1103/PhysRevLett.122.241605
[arXiv:1902.10571 [gr-qc]].

\end{thebibliography}

\end{document}